\documentclass[preprint,authoryear,12pt]{iopart}
\expandafter\let\csname equation*\endcsname\relax
\expandafter\let\csname endequation*\endcsname\relax
\usepackage{amsmath}
\usepackage{graphicx}
\usepackage{dcolumn}
\usepackage{bm}

\begin{document}

\title{\textit{Ab initio} parametrised model of strain-dependent solubility of H in $\alpha$-iron}
\author{D.~Psiachos}
\ead{psiachos@hubbard.physics.arizona.edu}
\address{Dept. of Physics, University of Arizona, Tucson Arizona 85721 USA}

\begin{abstract}
The calculated effects of interstitial hydrogen on the elastic properties of 
$\alpha$-iron from our earlier work are used to describe the H interactions with 
homogeneous strain fields using \textit{ab initio} methods. In particular we calculate the H solublility
in Fe subject to hydrostatic, uniaxial, and shear strain.
For comparison, these interactions are parametrised successfully
using a simple model with parameters entirely derived from \textit{ab initio} methods.
The results are used to predict the solubility of H in spatially-varying elastic strain fields, representative of
 realistic dislocations outside their core. 
We find a strong directional dependence of the H-dislocation interaction, leading to
strong attraction of H by the axial strain components of edge dislocations and by screw dislocations
oriented along the critical $\langle 111 \rangle$ slip direction. We further find a H concentration
enhancement around dislocation cores, consistent with experimental observations.
\end{abstract}


\maketitle

\section{Introduction}

The influx of hydrogen into many alloys and steels, either during manufacture
or service, is unavoidable and leads to undesirable consequences. One of these is the lowering of the failure stress, 
leading to embrittlement and fracture at unpredictable loading
conditions~\cite{Oriani,Hirth}. Some proposed mechanisms of H-embrittlement of iron include 
the hydrogen-enhanced decohesion (HEDE) mechanism~\cite{Troiano,Oriani2} in which H is postulated
to cause a loss of cohesion
between the metal atoms leading to failure at interfaces, 
H-vacancy effects~\cite{DawBaskes,Ohno,TateyamaPRB} in which vacancies containing H may order
themselves along critical slip directions,
and hydrogen-enhanced localised plasticity 
(HELP)~\cite{Beachem,BirnbaumMatSciEng,Robertson} where the onset of plasticity with 
loading occurs at a lowered stress
as a result of the H shielding of repulsive interactions between dislocations~\cite{BirnbaumMatSciEng}.
Due to the increased H concentration at dislocations~\cite{Cottrell}, this shielding
 effectively reduces
the spacing between dislocations~\cite{Robertson}. Support for the HELP hypothesis in explaining the
phenomenon of dislocation coalescence, and ultimately, crack advancement at
reduced loads, is provided by the experimentally-observed
increase of dislocation mobility~\cite{Robertson}. Various models of H-embrittlement
 are discussed in the context
of recent experimental results by Kirchheim~\cite{Kirchheim}.
H is very mobile in Fe, with a diffusivity of $10^{-8}-10^{-7}m^2/s$ as determined from a wide range
of experiments~\cite{hayashi}, and by calculations~\cite{JiangCarter}. Therefore,
 it can thus be assumed to be
present in all microstructural regions. In real materials with defects, H is most prevalent
at dislocations in the case of single crystal Fe, and at grain boundaries in polycrystalline Fe~\cite{Ono}.

The above theories assert that H degrades the material, but the details of the
mechanisms are still under debate~\cite{matsui,murakami}. Nevertheless, H-dislocation interactions have long 
been implicated in the modification (whether degradation or enhancement) of the strength properties
of iron~\cite{matsui,murakami} and recent experiments~\cite{Novak} confirm the importance 
of increased H-concentration near dislocations and other low-energy trap sites.

Compounding the confusion caused by the existence of multiple theories of 
H-embrittlement is that these various mechanisms have mainly been proposed and discussed separately and
rarely together~\cite{Gerberich, Novak} as would be desirable for an unbiased assessment of their
relative importance under various conditions.
The advantage of theroretical approaches is that the effects can be investigated independently.
However, a major challenge for achieving a theoretical understanding of 
these mechanisms lies in the vastly different length scales involved:
the large number of atoms required as a result of the slow spatial decay of 
dislocation strain fields and small H concentrations versus the
highly-localised modification of the electronic structure around H. As a result,
simulated systems need to be extremely large in order to account for the long-ranged 
elastic distortions induced by dislocations. Taketomi \textit{et al}~\cite{Taketomi2}
conducted a molecular-statics study with empirical potentials of the H distribution around an edge dislocation in 
Fe. Ref.~\cite{Rama} describes a kinetic Monte Carlo study using empirical potentials 
to examine diffusion and trapping of H in a bulk Fe crystal containing
screw dipoles. Clouet \textit{et al}~\cite{ClouetActaMat} have used empirical 
potentials combined with elasticity theory to model the interaction of C atoms with
edge and screw dislocations in Fe. However, empirical potentials have limited and
largely untested transferability to these systems~\cite{Rama2}, and more accurate studies 
using first-principles techniques are required. The current drawback with first-principles 
methods is that only small system sizes can be treated, leading to idealised
systems or conditions. The static calculation of the energy barrier for H-diffusion in bulk Fe~\cite{JiangCarter,Sanchez},
the study of the stress distribution around H in Fe~\cite{Sanchez}, or
a first-principles molecular-dynamics study of H-diffusion in a very small (16 atom) Fe unit cell~\cite{SanchezMD},
are typical of the types of problems that can currently be treated using first-principles methods.
Therefore, first-principles density functional theory is frequently combined with other techniques 
in order to examine effects requiring both atomistic accuracy and a large number 
of atoms. Examples of recent studies include a study of the microstructural
evolution of tungsten irradiated with He~\cite{BecquartHeW}, parametrising 
elasticity theory for studying solute-dislocation
interaction~\cite{sluiter}, and making extrapolations to continuum-scale properties such as changes in the shear modulus,
due to the effect of H on the electronic density of states of Fe~\cite{Gavriljuk}. 

The problem of interstitials interacting with dislocations has long been recognized to be important and has been studied analytically
~\cite{Cottrell,cochardt,eshelby1957,bilby,Hirth2,bockris,Zhang_strain} within continuum theory. A review of early work on the 
subject can be found in Ref.~\cite{fiore}. 
These studies recognized that the orientation of the dislocation with respect to the unit-cell 
axes can have dramatic effects on the
solubility and henceforth on the interstitial solute concentration around dislocation cores. 
However, while they are able to capture the effects of edge dislocations, whose 
strain fields are dominated by axial stresses, by not going beyond linear elasticity,
the models with hydrostatic expansion due to the solute fail to yield any interaction between solutes and screw 
dislocations, whose purely shear-stress fields
do not interact with the hydrostatic component of the strain field of the interstitial solute and studies thus ascribe the interaction of 
interstitials with screw dislocations purely to the tetragonal distortion which they introduce.
An earlier study~\cite{Zhang_strain} did
obtain interactions within linear elasticity theory with screw dislocations but only because
their model assumes a tetragonal distortion field generated by the interstitial solute. Very recently, the interaction
of C with dislocations was investigated by \textit{ab initio} methods~\cite{chrzan} in a model within linear
elasticity theory which accounted for tetragonal distortions.
This limitations of the models in properly describing
trapping of interstitials by screw dislocations has long been noted and 
ascribed to a non-linear elastic, or ``modulus effect" in the literature~\cite{eshelby1951,fiore,HirthLothe} but
studies of solute-dislocation interaction have mostly tended to ignore this effect. A recent exception is the work in
Ref.~\cite{wolfer} which studies the solute-dislocation interaction at the continuum scale.

The goal
of this study is to use \textit{ab initio} results from our previous study~\cite{psiachos1}
in order to make connections with the proposed theories of H-embrittlement for $\alpha$-Fe. Our 
results of the influence of interstitial H
on the elastic properties~\cite{psiachos1} are used to parametrise an elasticity-theory model 
for the behaviour of H within strain fields representative of dislocations and to predict
concentration enhancements around typical dislocation strain fields. 

In Sec.~\ref{sec:DFT} we describe the details of our DFT calculations used to calculate the solubility of H under various strain fields.
The elasticity-theory model of H solubility under spatially-homogeneous strain is 
introduced in Sec.~\ref{sec:strain-dep} and the predictions are compared
with the explicitly-calculated results. The model is further applied in Sec.~\ref{sec:conc} to study
the solubility in strain fields oriented in different directions with respect to the unit-cell axes
in order to establish a correspondence with locally-modified H concentrations in the vicinity of edge and screw dislocations. We 
conclude with a discussion in Sec.~\ref{sec:conclusions}.

\section{First-Principles Calculations}
\label{sec:DFT}
The details of the calculations 
are the same as in our previous study~\cite{psiachos1} and are therefore only summarised here.
Our spin-polarised first-principles density functional theory calculations were performed with 
the VASP~\cite{vasp1,vasp2,vasp3} code. 
We used the projector augmented-wave
method~\cite{paw1,paw2} and considered the 3p electrons of 
Fe as valence electrons. The generalized-gradient
approximation (GGA) in the PW91 parametrisation~\cite{pw91} was used for the exchange-correlation functional. 
A plane-wave basis with a cutoff of 500 eV
and supercell geometry were used. A $\Gamma$-centred k-point grid equivalent to 18$\times$18$\times$18 for the 
two-atom basis bcc unit cell was used for the Brillouin-zone sampling except for the 128-atom Fe cell where it was 
the equivalent of 20$\times$20$\times$20. 

 We focus on tetrahedral interstitial H because (i) at zero stress we find it to be 0.13 eV 
more stable than the octahedral position,
in agreement with previous DFT studies~\cite{JiangCarter,Sanchez}, and 
(ii) we expect it to dominate the mechanical properties at ambient
temperatures as there are only half as many octahedral interstitial sites. 
In all calculations, the cells with H were taken as cubic in the unstrained state, which
is a good approximation, for low concentrations of H, for
the determination of the elastic parameters~\cite{psiachos1}. The tetragonal distortion, which is usually accounted for
explicitly is studies of octahedral interstitials such as C in $\alpha$-Fe~\cite{Zhang_strain,chrzan}, is smaller
in the case of interstitials occupying tetrahedral sites. In addition, the cubic 
approximation makes sense as the occupancy of interstitial sites by H is expected to be random.

For the results presented here, we used different H concentrations which we 
arrived at using different supercells (Table II in Ref.~\cite{psiachos1}). For some of the cases studied (Sec.~\ref{sec:samestrs})
it was necessary to calculate the Fe and Fe-H cells at the same stress. Equal stresses in the two 
systems were maintained by enforcing lattice parameters such that the stresses in the
two systems are equal according to Eq.~\ref{eqstrs}, which 
is valid for small strains. The first-principles determination of stress,
from the DFT program, served as a further check that the systems had the
same stress.

\section{Strain dependence of H solubility}
\label{sec:strain-dep}
\subsection{Solution energy}
\label{sec:strain-dep1}
In this section, we describe the effect of different types of strain on H solubility. Our goal
is to examine the effect of H on typical dislocation strain fields and conversely, the latter's impact on H solubility.
Instead of simulating dislocations explicitly in computationally very demanding large
supercells, we have examined the solubility of H in 
different types of homogeneous strain fields which we later use to estimate the H solubility and henceforth the
concentration profile in realistic
dislocation strain fields.

Specifically, we have calculated the solution energy for tetrahedral interstitial 
H in supercells subject to hydrostatic, or uniaxial or shear 
strain. These applied strain tensors mimic the 
strain components of different dislocations as will be discussed in Sec.~\ref{sec:conc}. 

To arrive at the solution energy $W$ of $n$ H atoms, at each 
value of applied strain, we subtracted from the total energy
of the cell with $N$ Fe atoms and $n$ H atoms, $E_{Fe_NH_n}$, the total energy $E_{Fe_N}$ of a 
pure iron cell with $N$ Fe atoms, and half the energy of a free H$_2$ molecule multiplied by $n$: 
\begin{equation}
W=E_{Fe_NH_n}-E_{Fe_N}-n\frac{1}{2}E_{H_2}.
\label{solW}
\end{equation}

The solution energy can be interpreted in different ways depending on the strain or stress state 
of the pure Fe reference system with respect to the system with H. We have defined three regimes on 
which to concentrate as the strain of the Fe-H system is varied:
(i) the pure Fe volume matching that of the Fe-H system at each strain, (ii) the two systems
having the same stress tensor at each strain, or (iii) having the same strain tensors. Cases (i) and
(iii) would be described by the change in Helmholtz free energy with constant volume and temperature, and
constant strain and temperature respectively, with H, while case (ii) would be described by the change in the Gibbs free energy
where temperature and pressure are maintained steady. Case (i)
would reflect the energy to be gained or lost by inserting $\frac{1}{2}H_2$ into a strained Fe lattice
without further changing the average volume per Fe atom. It would be the expected boundary condition for H in an perfect infinite Fe lattice. It would also
correspond to regions where H is present in already-expanded regions of the metal, \textit{e.g.} near a grain
boundary or a dislocation. Cases (ii) and (iii) would involve other types of confinement, for instance, in the case of H locally present at crack tips, the applied boundary conditions would determine
whether any eventual failure is stress- or strain-controlled. The boundary conditions chosen are important for a variety of systems,
affecting, for example, the microstructure of alloys~\cite{LiChen1} or the activation energy of dislocation nucleation~\cite{Ryu}.

\subsection{Parametrisation of the H solution energy}
Using our DFT-obtained material parameters (Table~\ref{params}), we have parametrised 
the solution energies (Eq.~\ref{solW}) as functions of strain and H-concentration. 
We label all quantities pertaining to the Fe-H or the pure Fe system with an $x$ or $0$ superscript respectively:
 for example, $V^Y$ for the equilibrium volume where $Y=x$ for the Fe-H system
and $Y=0$ for pure Fe.
For a system $Y$ the total energy at an arbitrary reference 
volume $V^{ref}$ corresponding to a stress tensor $\boldsymbol{\sigma}^{ref}$ 
and infinitesimal strain tensor $\boldsymbol{\epsilon}$ with respect to $V^{ref}$ is (\textit{c.f.} Eq. 6 in
Ref.~\cite{psiachos1}) comprised of three parts: a reference term, a stress term, and a strain term,
\begin{equation}
E^{Y}\left(V^{ref},\boldsymbol{\epsilon}\right)=E^{Y}(V^{ref},0)
+V^{ref}\sum_{ij}\sigma^{ref}_{ij}\epsilon_{ij}
+\frac{V^{ref}}{2}\sum_{ijkl}C^Y_{ijkl}(V^{ref})\epsilon_{ij}\epsilon_{kl}.
\label{Esum}
\end{equation}
The first term is the total energy at the reference volume $V^{ref}$, the second term is the stress part 
which vanishes for $V^{ref}=V^Y$,
and the last term is the elastic deformation energy about $V^{ref}$. The expression~(\ref{Esum})
equivalently arises from an expansion of the elastic energy from $V^Y$ up to third order in strain defined
with respect to $V^Y$. 
Explicitly, the reference term in Eq.~\ref{Esum} is equal to
\begin{eqnarray}
E^Y(V^{ref},0)&=& E^Y(V^Y,0) + \frac{1}{2}\left(V^{ref}-V^Y\right)^2\left.\frac{\partial^2 E^Y}{\partial V^2}\right|_{V=V^Y}\nonumber\\
&=&E^Y(V^Y,0)+\frac{B^Y}{2V^Y}\left(V^{ref}-V^Y\right)^2,
\label{eqnelecFe}
\end{eqnarray}
where $B^Y$ is the bulk modulus of system $Y$ at its equlibrium volume $V^Y$. The 
second term in Eq.~\ref{Esum}, assuming a hydrostatic stress corresponding to 
a pressure $P^{ref}$ required to reach the reference volume, is
\begin{eqnarray}
V^{ref}\sum_{ij}\sigma^{ref}_{ij}\epsilon_{ij}&=&-P^{ref}V^{ref}\sum_{ij}\delta_{ij}{\epsilon}_{ij}\nonumber\\
&=&-P^{ref}V^{ref}\sum_{i}\epsilon_{ii}\nonumber\\
&=&\frac{B^YV^{ref}}{V^Y}\left(V^{ref}-V^Y\right)\sum_{i}\epsilon_{ii},
\label{pressure}
\end{eqnarray}
where $P^{ref}=-B^Y\left(V^{ref}-V^Y\right)/V^Y$ was used. This relation for $P^{ref}$ makes
use of linear elasticity, \textit{i.e.} that $B^Y$ doesn't vary, in Eq.~\ref{pressure}, 
which is a reasonable assumption for low concentrations. 
 In the third term of Eq.~\ref{Esum}, the 
elastic parameters at the volume $V^{ref}$ are equal to
\begin{eqnarray}
C_{ijkl}^Y(V^{ref})&=&C_{ijkl}^Y(V^Y)+\left(V^{ref}-V^Y\right)\left.\frac{\partial C_{ijkl}^Y}{\partial V}\right|_{V=V^Y}\nonumber\\
&=&C_{ijkl}^Y(V^Y)+\frac{V^{ref}-V^Y}{V^Y}\left.\frac{\partial C_{ijkl}^Y}{\partial \eta}\right|_{V=V^Y},
\label{conc}
\end{eqnarray}
where the volume derivative has been rewritten in terms of the 
hydrostatic strain $\eta\left(V\right)\equiv \left(V-V^Y\right)/V^Y$.

Despite their generality, Eqs.~\ref{eqnelecFe}-\ref{conc} are difficult to 
work with for the Fe-H system, due to the complex strain-dependency of 
the $C_{ijkl}^Y$. We prefer therefore to express all quantities in terms of small variations from $x=0$ and $V^0$, and we
take the reference volume as $V^{ref}=V^x$ and expand $V^x$ about $V^0$ as
$V^x=V^0+xdV/dx$ (for small $x$). With these expansions, we arrive at a general form
of Eq.~\ref{Esum} for the Fe-H system that we will use throughout the applications presented
in the following sections. The first term of Eq.~\ref{Esum} (\textit{i.e.} Eq.~\ref{eqnelecFe}) is
thus equal to 
\begin{eqnarray}
E^x(V^x,0)&=&E^0(V^0,0)+x\left.\frac{\partial E}{\partial x}\right|_{V^0,x=0}
+x\left(V^x-V^0\right)\left.\frac{\partial^2 E}{\partial x \partial V}\right|_{V^0,x=0}\nonumber\\
&\qquad&+\frac{1}{2}\left(V^x-V^0\right)^2\left.\frac{\partial^2 E}{\partial V^2}\right|_{V^0,x=0}+\frac{1}{2}x^2\left.\frac{\partial^2 E}{\partial x^2}\right|_{V^0,x=0}\label{e_elec2}\\
&=&E^0(V^0,0)+x\frac{\partial E}{\partial x}+x^2\frac{dV}{dx}\frac{\partial^2 E}{\partial x \partial V}+\frac{B^0}{2V^0}\left(x\frac{dV}{dx}\right)^2+\frac{1}{2}x^2\frac{\partial^2 E}{\partial x^2}\nonumber,
\end{eqnarray}
where here and in what follows, the limits of evaluation of the derivatives are omitted for clarity.
The first derivative with respect to volume vanishes as the evaluation limits correspond to pure Fe. 
The second term of Eq.~\ref{Esum} (\textit{i.e.} Eq.~\ref{pressure}) also vanishes because 
$V^x$ is the system's equilibrium volume. For the third term, the elastic
parameters (Eq.~\ref{conc}) can be written as
\begin{equation}
C_{ijkl}^{x}\equiv C_{ijkl}^{x}(V^x)=C_{ijkl}^0+x\frac{dC^{tot}_{ijkl}}{dx},
\label{conc2}
\end{equation}
where the derivative represents the variation with respect to 
concentration about $V^0$ and with strain included implicitly. These 
derivatives were determined in Ref.~\cite{psiachos1}
and are given in Table~\ref{params}.

\begin{table}[htb]
\caption{\label{params}
Parameters used in the solubility expressions~\ref{sol3bcub}-\ref{hydroeqn2} and in subsequent sections.
All derivatives are evaluated at $V^0$ and $x=0$.} 
\begin{indented}
\item[]\begin{tabular}{|c|c|}\hline
Quantity&value\\
\hline
$\Delta$ E&0.22 eV\\
$B^0$&194.2 GPa\\
$\Delta$ V&4.5 \AA$^3$\\
$V^0/N$&11.352 \AA$^3$\\
$dB^0/d\eta$&-1075 GPa\\
$dC^0_{11}/d\eta$&-1492 GPa\\
$dC^0_{12}/d\eta$&-669 GPa\\
$dC^0_{44}/d\eta$&-761 GPa\\
$dB^{tot}/dx$&-153 GPa\\
$dC_{11}^{tot}/dx$&-316 GPa\\
$dC_{12}^{tot}/dx$&-72 GPa\\
$dC_{44}^{tot}/dx$&-141 GPa\\
\hline
\end{tabular}
\end{indented}
\end{table}

After defining the relative volumes, stresses, and/or strains, of the Fe-H and Fe systems corresponding to
each of the cases outlined above, we use Eqs.~\ref{solW}
and~\ref{Esum} to derive strain-dependent solution energies for different 
relative strain and/or volume conditions for the Fe and Fe-H systems.

\vspace{3mm}
\subsection{Applications of solution energy model to limiting cases}
\subsubsection{Same volumes}
\label{samevols}
The reference volume of the Fe-H system, $V^x$, is taken as the reference volume 
of the pure Fe system, resulting in a contribution from the stress term (Eq.~\ref{pressure}).
The pure Fe system, at a reference volume $V^x$ and strain $\epsilon^x$, 
applying the equations of the previous 
section (Eq.~\ref{Esum}-\ref{pressure}), 
is thus described by a total energy
\begin{equation}
E^{0}\left(V^x,\boldsymbol{\epsilon}^x\right)=E^{0}\left(V^x,0\right)+\frac{B^0}{2V^0}\left(x\frac{dV}{dx}\right)^2+B^0\frac{V^x}{V^0}x\frac{dV}{dx}\sum_i\epsilon_{ii}^x+\frac{V^x}{2}\sum_{ijkl}C^0_{ijkl}(V^x)\epsilon^x_{ij}\epsilon^x_{kl}
\end{equation}
where the elastic parameters are (Eq.~\ref{conc}), 
\begin{eqnarray}
C^0_{ijkl}(V^x)&=&C^0_{ijkl}\left(V^0\right)+\frac{V^x-V^0}{V^0}\frac{\partial C_{ijkl}^0}{\partial \eta}\nonumber\\
&=&C^0_{ijkl}\left(V^0\right)+\frac{x}{V^0}\frac{dV}{dx}\frac{\partial C_{ijkl}^0}{\partial \eta}.
\end{eqnarray}

The solution energy (Eq.~\ref{solW}) for a cubic system is then given by the 
contributions to Eq.~\ref{Esum} where we
 neglect all terms of order $x^2$ except for those in Eq.~\ref{e_elec2}, which are required for 
explaining the discrepancies at zero strain between the various concentrations:
\begin{eqnarray}
W/n&=& \Delta E  +\frac{x}{2N}\frac{\partial^2E}{\partial x^2}+x\Delta V \frac{\partial^2 E}{\partial x \partial V}
-B^0\Delta V \left(\epsilon^x_{xx}+\epsilon^x_{yy}+\epsilon^x_{zz}\right)\nonumber\\
+&\frac{1}{2}&\left[\left(\frac{V^0}{N}\frac{dC_{11}^{tot}}{dx}-\frac{dC^0_{11}}{d\eta}\Delta V \right)\left({\epsilon^x_{xx}}^2+{\epsilon^x_{yy}}^2+{\epsilon^x_{zz}}^2\right)\right.\nonumber\\
&+&2\left(\frac{V^0}{N}\frac{dC_{12}^{tot}}{dx}-\frac{dC^0_{12}}{d\eta}\Delta V \right)\left(\epsilon^x_{xx}\epsilon^x_{yy}+\epsilon^x_{xx}\epsilon^x_{zz}+\epsilon^x_{yy}\epsilon^x_{zz}\right)\nonumber\\
&+&\left.4\left(\frac{V^0}{N}\frac{dC_{44}^{tot}}{dx}-\frac{dC^0_{44}}{d\eta}\Delta V \right)\left({\epsilon^x_{xy}}^2+{\epsilon^x_{xz}}^2+{\epsilon^x_{yz}}^2\right)\right].
\label{sol3bcub}
\end{eqnarray}
In the above, we have used the definition of concentration as the ratio of the 
number of H atoms, $n$, to the number of Fe atoms $N$, $x=n/N$,
to rewrite the concentration derivative of $V$ in terms of the volume expansion $\Delta V$ per H and $N$: 
$\frac{dV}{dx}=\frac{dV}{dn}\frac{dn}{dx}={\Delta V}{N}$ and similarly for the concentration derivative of $E$,
$\frac{dE}{dx}={\Delta E}{N}$ where $\Delta E=\frac{dE}{dn}-\frac{1}{2}E_{H_2}.$  The 
ratio $V^0/N$ is the (constant) volume per Fe atom at equilibrium. In Eq.~\ref{sol3bcub}, the $x$-dependent
terms - $O(x^2)$ in $W$ - are given only to highlight the higher-order effects at zero strain; otherwise
the expression constitutes a universal formula valid for all low concentrations.
We obtained a value of $\Delta E$=0.22 eV by calculating the $\frac{1}{2}H_2$ solution energy in the
largest unit cell (128 atoms) studied. 
The expression for the solution energy can be simplified for particular cases of applied strain.  
For the case of hydrostatic strain $\epsilon=\epsilon_{xx}=\epsilon_{yy}=\epsilon_{zz}$, 
Eq.~\ref{sol3bcub} becomes
\begin{eqnarray}
{W}^{hydro}/n=\Delta E&+&\frac{x}{2N}\frac{\partial^2E}{\partial x^2}+x\Delta V\frac{\partial^2 E}{\partial x \partial V}-3B^0\Delta V \epsilon\nonumber\\
&+&\frac{9}{2}\left(\frac{V^0}{N}\frac{dB^{tot}}{dx}-\frac{1}{3}\frac{d\left(C_{11}^0+2C_{12}^0\right)}{d\eta}\Delta V\right)\epsilon^2.
\label{hydroeqn}
\end{eqnarray}
Here, we have not associated the quantity $\frac{1}{3}\frac{d\left(C_{11}^0+2C_{12}^0\right)}{d\eta}$ with 
$\frac{dB^0}{d\eta}$ because the slopes of the energy-strain coefficients as a function of pressure are not equal to those of the stress-strain 
coefficients~\cite{barron,Wallace} used to calculate the bulk modulus. However, the 
numerical values of the respective slopes are not very different~\cite{psiachos1}. The simplified, linear in strain, form of Eq.~\ref{hydroeqn},
$W/n=\Delta E-3B^0\Delta V \epsilon$, is a well-known result of the literature of continuum mechanics~\cite{Cottrell,eshelbySSP}.

The corresponding expressions for uniaxial strain $\epsilon\equiv \epsilon_{xx}$ 
and shear strain $\epsilon\equiv \epsilon_{xy}=\epsilon_{yx}$ are
\begin{equation}
{W}^{uniax.}/n=\Delta E+\frac{x}{2N}\frac{\partial^2E}{\partial x^2}+x\Delta V \frac{\partial^2 E}{\partial x \partial V}-B^0\Delta V\epsilon
+\frac{1}{2}\left(\frac{V^0}{N}\frac{dC_{11}^{tot}}{dx}-\frac{dC_{11}^0}{d\eta}\Delta V\right)\epsilon^2
\label{uniaxeqn}
\end{equation}
\begin{equation}
{W}^{shear}/n=\Delta E+\frac{x}{2N}\frac{\partial^2E}{\partial x^2}+x\Delta V\frac{\partial^2 E}{\partial x \partial V}+2\left(\frac{V^0}{N}\frac{dC_{44}^{tot}}{dx}-\frac{dC_{44}^0}{d\eta}\Delta V \right)\epsilon^2.
\label{pshreqn}
\end{equation}
A combination of hydrostatic/uniaxial/shear strain corresponds to combining expressions~\ref{hydroeqn}-\ref{pshreqn}
as can be seen from Eq.~\ref{sol3bcub}.

\begin{figure}
\includegraphics[angle=0,width=4.5in]{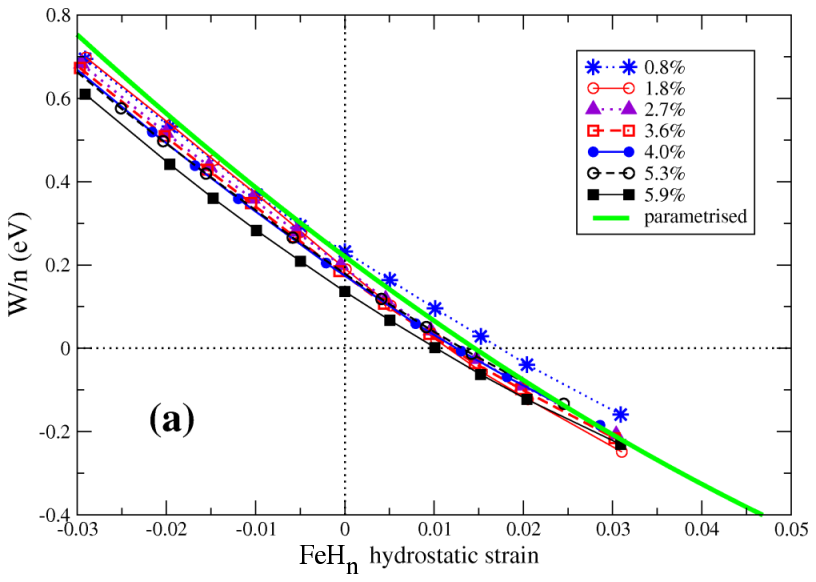}
\includegraphics[angle=0,width=4.5in]{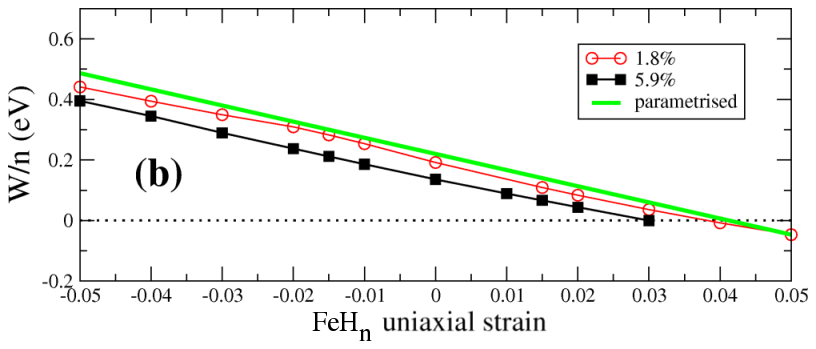}
\includegraphics[angle=0,width=4.5in]{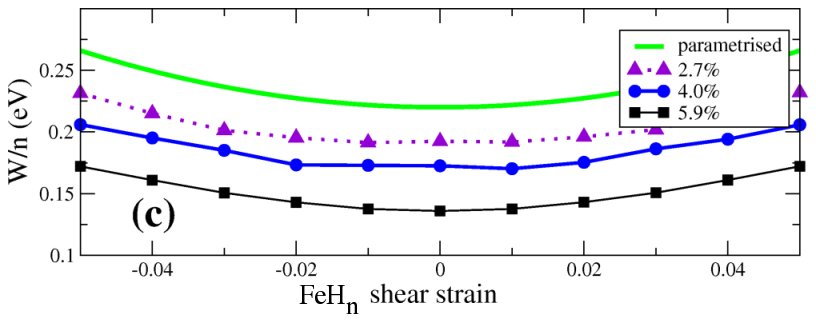}
\caption{[color online] DFT-calculated solution energies along with parametrised form for (a) hydrostatic strain,
(b) uniaxial strain, (c) shear strain, for different H-concentrations. All strains are component (length) strains.}
\label{solE}
\end{figure}

By evaluating the coefficients of Eq.~\ref{sol3bcub}, apart from those of the $x$-dependent terms 
which are discussed below, using the values given in Tab.~\ref{params}, we obtain
a parametrisation of the H solution energy for small arbitrary strain, pressure, 
and H-concentration.  To that end, we performed DFT simulations with the Fe and Fe-H systems at 
the same volume and varied the amount
and type of strain. We subtracted the two sets of energies and the energy of $\frac{1}{2}H_2$ (Eq.~\ref{solW}).
The parametrised forms (Eqs.~\ref{hydroeqn}-\ref{pshreqn}), using the DFT-calculated parameters 
listed in Table~\ref{params}, accurately describe the DFT-obtained solution energies, 
as can be seen from Figure~\ref{solE}. The parametrised curves in all of the figures were generated
by assuming the dilute limit $\left [\lim_{x\to 0}\left(\frac{W}{n}\right)\right ]$ of Eq.~\ref{sol3bcub}, 
such that the contribution from the $x$-dependent parts of Eqs.~\ref{sol3bcub}-\ref{hydroeqn2} is zero.

The variations of the calculated solution energies at zero strain among the various concentrations can
be accounted for by evaluating the terms 
$x\frac{\partial E}{\partial x}+x^2\frac{dV}{dx}\frac{\partial^2 E}{\partial x \partial V}+\frac{1}{2}x^2\frac{\partial^2 E}{\partial x^2}$ 
(\textit{c.f.} Eq.~\ref{e_elec2} and the normalised form in Eq.~\ref{sol3bcub}) explicitly at each concentration
as summarised in Tab.~\ref{crossterm}.
The cross-term $\frac{\partial^2 E}{\partial x \partial V}$ was calculated for $x$=1.8\% and this 
result (-0.557 eV/\AA) was used throughout. 
Using a graph similar to Fig.~\ref{solE}a, but with the horizontal axis expressed in terms of the strain $\epsilon^0$ of pure Fe, 
the sum of the first two terms of Eq.~\ref{sol3bcub} was predicted
using the zero-strain data (equal to a volume of $V^0$) from there. This can be done
since $\left. W(x)\right|_{\epsilon^0=0}=\Delta E+\frac{x}{2N}\frac{d^2 E}{dx^2}.$
These sums are given as the second column of Table~\ref{crossterm}. The total sums including the cross-term, stated in the fourth column, agree very well with the DFT data
given in the last column.
The cross-term was also calculated using data from the 5.9\% concentration and it was a bit smaller (-0.426 eV/\AA)
than the one used for Table~\ref{crossterm}, leading to a contribution $x\Delta V\frac{\partial^2 E}{\partial x \partial V}$=-0.114 eV and thus a total
$E^x(V^x,0)$=0.141 eV, for the entry `sum' at $x=5.9\%$, thus accounting for the minor deviations at the higher concentrations between
the two last columns of the table. The deviation of the intercept from the zero-concentration limit of 0.22 eV 
for all the graphs shown in Fig.~\ref{solE} can thus be explained.

\begin{table}[htb] 
\caption{\label{crossterm} Energy terms from Eq.~\ref{eqnelecFe} contributing to the intercept at zero strain
 for the case of equal volumes (Fig.~\ref{solE}).
 The first two energy columns are from expression~\ref{e_elec2} and were calculated using DFT 
as described in the text. Their sum is under the heading \textit{sum}. The last column
is the DFT data.}
\begin{indented}
\item[]\begin{tabular}{|c||c|c|c|c|}\hline
 $x(\%)$&$\Delta E+\frac{x}{2N}\frac{d^2 E}{dx^2}$(eV)&$x\Delta V\frac{\partial^2 E}{\partial x \partial V}$(eV)&sum&DFT $\left. W\right|_{\mathbf{\epsilon^x}=0}$(eV)\\
\hline
0.8&0.246&-0.023&0.223&0.229\\
1.8&0.238&-0.048&0.190&0.192\\
2.7&0.249&-0.066&0.183&0.190\\
3.6&0.251&-0.089&0.162&0.175\\
4.0&0.250&-0.093&0.157&0.173\\
5.3&0.278&-0.123&0.155&0.177\\
5.9&0.255&-0.149&0.106&0.136\\
\hline    
\end{tabular} 
\end{indented}
\end{table}         

The solution energy as a function of 
strain and concentration can also be derived by a systematic expansion of the total energy in terms of strain and concentration~\cite{sluiter}. In that work, the authors
presented a systematic expansion of the total energy in terms of strain and concentration. 
The resulting derivative
of pressure with respect to concentration yields a contribution to the solution energy which is linear in hydrostatic
strain, in agreement with our findings, and with the earlier work in Refs.~\cite{Cottrell,eshelbySSP} which assumed
a continuum model throughout. Using the notation of Ref.~\cite{sluiter} to express our concentration and strain-dependent elastic parameters
requires considering up to the third derivatives 
of the total energy with respect to concentration and strain but otherwise yields equivalent results.

\subsubsection{Same stresses or same strains}
\label{sec:samestrs}
As discussed at the end of Sec.~\ref{sec:strain-dep1}, another possibility of evaluating the solution energy (Eq.~\ref{solW}) 
is to compare the Fe-H and pure Fe 
reference systems for the same stress or strain tensors. Therefore, we expand the total energy 
of each system about its equilibrium volume.

Using Eq.~\ref{Esum}, the same stress $\sigma_n$, for the two systems corresponds to 
\begin{equation}
\sigma_n=\sum_{m} C^0_{nm}\epsilon^0_m=\sum_k C^x_{nk}\epsilon^x_k
\label{eqstrs}
\end{equation}
in the Voigt notation for m,n,k from 1 to 6. In the case of hydrostatic 
strain $\epsilon^x_{ij}=\delta_{ij}\epsilon$, 
this leads to $\epsilon^0_{ii}=\epsilon\frac{B^x}{B^0}$. An expansion of the stress (energy) to
second (third)-order in strain leads to contributions in the solution energy of 
powers of $\epsilon^3$ which we ignore.
If instead of equal stresses, the systems have 
equal strain tensors, the relation is simply $\epsilon^0_{ij}=\epsilon^x_{ij}.$

We obtain, as a function of Fe-H hydrostatic strain $\epsilon^x_{ij}=\delta_{ij}\epsilon$
 and shear strain $\epsilon^x_{xy}=\epsilon^x_{yx}=\delta$, the following expression
for the solution energy per H:
\begin{eqnarray}
W/n=\Delta E&+&\frac{x}{2N}\frac{\partial^2E}{\partial x^2}+x\Delta V \frac{\partial^2 E}{\partial x \partial V}+\frac{B^0}{2}\frac{N}{V^0}x\Delta V^2\nonumber\\
&+&\hspace{-1mm}\frac{9}{2}\left[B^0\Delta V \pm \frac{V^0}{N}\frac{dB^{tot}}{dx}\right]\epsilon^2+2\left[C_{44}^0\Delta V \pm\frac{V^0}{N}\frac{dC_{44}^{tot}}{dx}\right]\delta^2.
\label{hydroeqn2}
\end{eqnarray}
\begin{figure}
\includegraphics[angle=0,width=4.5in]{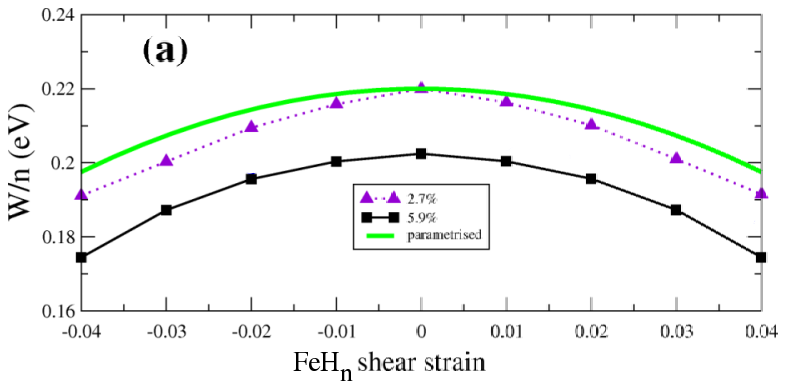}
\includegraphics[angle=0,width=4.5in]{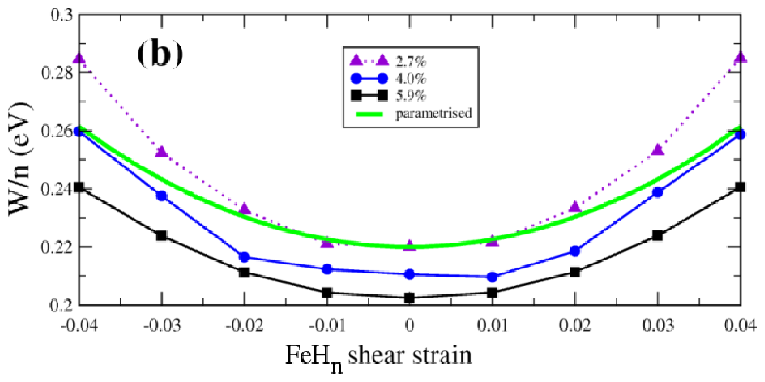}
\caption{[color online] DFT-calculated solution energies per H atom for various concentrations 
along with parametrised forms (Eq.~\ref{hydroeqn2})
 as a function of shear strain where the reference pure Fe system has (a) the same strains, 
and (b) the same stresses. The strain on the horizontal axis is for the Fe-H system.}
\label{solEstress}
\end{figure}

The upper (lower) signs are for equal strains (stresses). Eq.~\ref{hydroeqn2}
is sufficiently general such that expressions can be found for mixed cases, \textit{e.g.} equal hydrostatic
stresses but equal shear strains.
The expression~\ref{hydroeqn2} is compared with directly explicit DFT calculations in Figure~\ref{solEstress}
for $\epsilon=0$ and varying $\delta$ for the cases of (a): equal strains, and (b): equal stresses.
The higher
curvature for $x$=2.7\% in (b) is due to the $C_{44}$
data (Fig. 4 in Ref.~\cite{psiachos1}) lying below the line of best fit, 
thereby yielding a larger contribution from the total derivative $dC^{tot}_{44}/dx$.
\begin{table}[htb]
\caption{\label{crossterm2} Energy terms contributing to the intercept at zero strain (Fig.~\ref{solEstress})
for the case of equal strains or stresses. The first energy column is the same as that in Table~\ref{crossterm}
while there is now an additional term (second energy column in this table) appearing.
The sum of the first two columns is under
the third column, \textit{sum}, which should be compared with the DFT data in the last column.}
\begin{indented}
\item[]\begin{tabular}{|c||c|c|c|c|}\hline
$x(\%)$&Table~\ref{crossterm} sum&$\frac{B^0}{2}\frac{N}{V^0}x\Delta V^2$&sum&DFT $\left. W(x)\right|_{\mathbf{\epsilon^x}=0}$(eV)\\
\hline
2.7&0.183&0.030&0.213&0.220\\
4.0&0.157&0.045&0.202&0.211\\
5.9&0.106&0.068&0.174&0.202\\
\hline
\end{tabular}
\end{indented}
\end{table}

As with the previous case of equal volumes, the intercept at zero strain in Fig.~\ref{solEstress} can be accounted for by explicitly calculating
the zero-strain terms appearing in Eq.~\ref{hydroeqn2}, along with an extra term (third column of Table~\ref{crossterm2}). 
The comparison with the DFT data, in Table~\ref{crossterm2},
is very good for small concentrations and deviates for the larger one, by a similar amount as the case of constant volumes.
As in that case, discussed at the end of Sec.~\ref{samevols}, the deviation can be attributed
 to variation of the cross-term with concentration.

\section{H solubilities in realistic dislocation strain fields}
\label{sec:conc}
\subsection{Anisotropy of H solubility}
The parametrisation of the H solution energy in terms of strain presented in the previous
section enables us to determine the H solubility in typical dislocation strain fields. 
We will focus on the particular case of a screw dislocation lying along $\langle 111 \rangle$, which
is the predominant orientation of a screw dislocation in a bcc metal~\cite{screw1}.
Within isotropic dislocation theory~\cite{HirthLothe}, the strain field of a screw dislocation
is described in cylindrical $(r,\theta,z)$ coordinates by its
only nonzero component as
\begin{equation}
\epsilon_{\theta z}=\epsilon_{z\theta}=\frac{b}{4\pi r}
\label{burgers}
\end{equation}
 where
$b=2.45$~\AA~ is a Burgers vector equal to the length of $\frac{1}{2}\langle 111 \rangle$ for a dislocation
lying along $\langle 111 \rangle$.

The solution energy for pure shear strain at the same volume (Eq.~\ref{pshreqn}) and 
at the same stress/strain (Eq.~\ref{hydroeqn2}) is independent of the choice of shear direction as 
long as it is oriented along the cube axes \textit{i.e.} if 
$\epsilon^x_{xy}=\epsilon^x_{yx}=\delta$, $\epsilon^x_{xz}=\epsilon^x_{zx}=\delta$, or 
$\epsilon^x_{yz}=\epsilon^x_{zy}=\delta$. A screw dislocation line oriented along $\langle 001\rangle$ 
would result in shear components oriented along the cube axes, specifically with 
$\epsilon^x_{xz}=\epsilon^x_{zx}\neq 0$, $\epsilon^x_{yz}=\epsilon^x_{zy}\neq 0$, 
and $\epsilon^x_{xy}=\epsilon^x_{yx}=0$.
Referring to Fig.~\ref{solEstress}, we see that
H is attracted towards the case of the screw dislocation in the case where the pure Fe reference 
system has the same shear strain as the Fe-H system (Fig.~\ref{solEstress}a), but repelled in the
case of same stresses (Fig.~\ref{solEstress}b) and same volume (Fig.~\ref{solE}). 

However, such an attractive behaviour can vary significantly with the direction of applied 
shear strain~\cite{Taketomi3}. Therefore, the solubility equations~\ref{hydroeqn}-\ref{pshreqn} and \ref{hydroeqn2} 
were transformed to rotated coordinates corresponding to orientations other than the 
$\langle 100 \rangle$-type considered in previous sections. 
By rotating the entire coordinate system,
$(x,y,z)\rightarrow\left(x^\prime,y^\prime,z^\prime\right)$, other directions $z^\prime$
for the dislocation line are achieved.
The transformation of strains from the reference (unprimed)
to rotated (primed) coordinate system can be performed using Euler angles (see \textit{e.g.} Ref.~\cite{HirthLothe}):
$\boldsymbol{\epsilon}^\prime=\mathbf{T}\boldsymbol{\epsilon}\mathbf{T}^\mathrm{T}$ where $\mathbf{T}$ is the transformation matrix for
transforming the reference coordinates $\mathbf{x}$ into the new frame $\mathbf{x^\prime}$,
such that $\mathbf{x}^\prime=\mathbf{T}\mathbf{x}$.
We expressed the strains 
$\epsilon_{x^\prime z^\prime}=\epsilon_{z^\prime x^\prime}$ in the
primed system in terms of the unprimed system and inserted these
values into the general expression for the solution energy Eq.~\ref{sol3bcub} for the case
of equal volumes and its equivalent general form for the cases of equal strains/stresses. 

\begin{figure}
\includegraphics[angle=0,width=5.in]{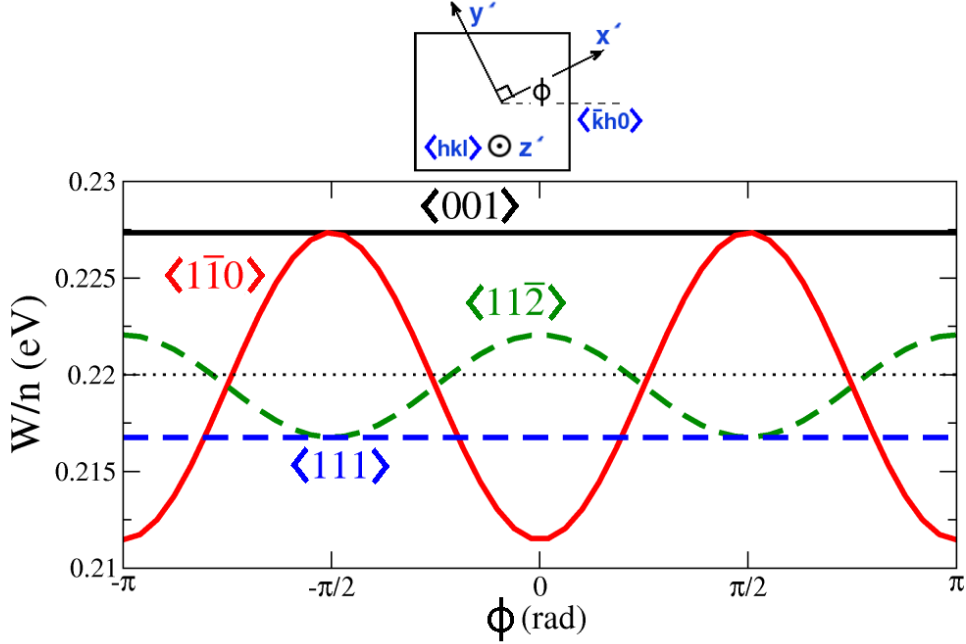}
\caption{[color online] Variation of solution energy for 2\% shear strain $\epsilon_{y^\prime z^\prime}$ (see inset) in 
the case of 
equal volumes for a hypothetical screw dislocation line/Burgers vector oriented
along different $z^\prime$ directions (labels and $\langle hkl \rangle$ in inset). 
The horizontal axis is the variation of the solution energy as a 
function of the angle $\phi$ which represents the choice of $x^\prime$ and $y^\prime$ axes within the plane
normal to the Burgers vector. $\phi=0$ has been arbitrarily chosen to correspond
to the $x^\prime=\langle\overline{k}h0 \rangle$ direction 
(see inset). The dotted line is the unstrained value of the H solution energy.}
\label{solscrew}
\end{figure}

For the case of purely hydrostatic strain, the solution energy is independent 
of the coordinate system chosen because
the trace of the strain tensor is invariant under any unitary transformation. Figure~\ref{solscrew} 
shows the variation of the solution energy in the case of equal volumes, 
for $\epsilon_{y^\prime z^\prime}$=2\% where
Eq.~\ref{sol3bcub} has been rotated onto other axes. These axes are shown in the inset where the dislocation is taken
to lie along $z^\prime$. The $z^\prime$ direction may be associated with the Miller indices $hkl$ of a plane.
$\phi$, the angle that $x^\prime$ makes with the $\langle\overline{k}h0 \rangle$ direction, also defines the direction of $y^\prime$.
In Fig.~~\ref{solscrew} only the 
$\epsilon^x_{y^\prime z^\prime}=\epsilon^x_{z^\prime y^\prime}$ components
are applied. For the full description of the screw dislocation strain field $\epsilon_{\theta^\prime z^\prime}$, the 
$\epsilon^x_{x^\prime z^\prime}=\epsilon^x_{z^\prime x^\prime}$ components are
also required. In the case of the $\langle 111 \rangle$ screw dislocation line, the solution energy as a function
of the full strain field $\epsilon_{\theta^\prime z^\prime}$ remains flat, independent of shear direction $\phi$.
In contrast with Ref.~\cite{Taketomi3},
we do not find asymmetry of our solution energy $W$ about zero strain as a function of applied shear strain, but
we cannot provide any insight into this discrepancy 
because the details of the model used in Ref.~\cite{Taketomi3} are lacking.

\begin{figure}
\includegraphics[angle=0,width=5.in]{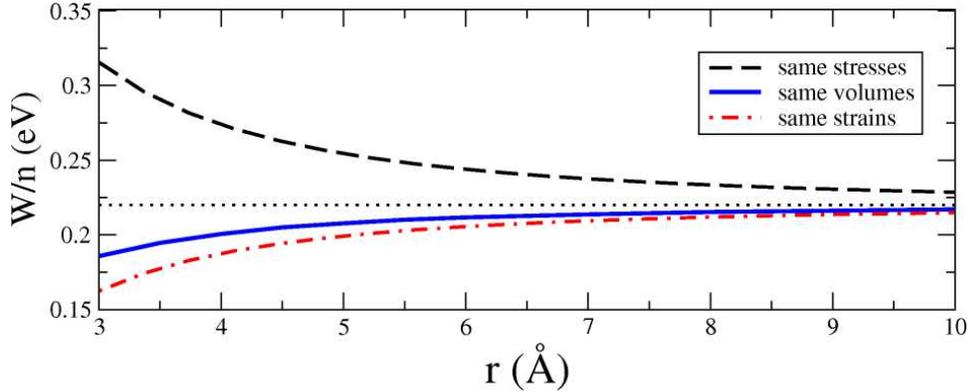}
\caption{[color online] Variation of H solubility energy as a function of distance from the $\langle 111 \rangle$ screw dislocation core
for the cases of equal stresses, equal volume, or equal strains for the Fe-H and pure Fe systems. The dotted
line denotes the zero-strain solubility. The strain field of the screw dislocation was represented by Eq.~\ref{burgers}.}
\label{solscrewdisl}
\end{figure}

The curves in Fig.~\ref{solscrew}, while corresponding to the same 
magnitude of shear, do not necessarily correspond to the same
distance $r$ away from the dislocation core, as the Burgers vector (\textit{c.f.} Eq.~\ref{burgers}) is different for dislocation lines
lying along different directions. The radial dependence of the H solubility was calculated for the case 
of a $\langle 111\rangle$ dislocation for the three cases
described in Sec.~\ref{sec:strain-dep}. The solubility corresponding to
the complete strain tensor Eq.~\ref{burgers} was calculated from Eq.~\ref{sol3bcub} in the case of equal volumes or analogous
equations for the
cases of equal stresses and equal strain. Regarding the case of equal stresses, we have not modelled 
the solubility of H in the stress field of a dislocation \textit{per se} but 
its solubility at individual stresses at each point $r$, meaning that
each infinitesimal annular element is not in mechanical equilibrium with its surroundings. 
However, this is a first approximation in a self-consistent treatment of H solubility - see
\textit{e.g.} Ref.~\cite{SofronisJMechPhysSol}.
The results, shown in Fig.~\ref{solscrewdisl}, show that H 
is attracted to the core for equal shear strains or equal volumes
and repelled for equal shear stresses.

\subsection{Concentration of H around dislocation strain fields}
Motivated by the strong strain-solubility effects, we calculated the local modification of the
H concentration by strain fields for the case of equal volumes.
We consider the change in free energy $F$ as a result of adding H to Fe under strain. The free energy is
\begin{equation}
F=E_{Fe-H}-TS,
\label{omega}
\end{equation}
where $T$ is the temperature, and $S$ is the total entropy of the Fe-H system. In order to isolate the effect
of adding H to the already-strained Fe lattice, we assume that the configurational 
entropy of the H is independent of that
of the Fe lattice: \textit{i.e.} $S=S_{H}+S_{Fe}$. $E_{Fe-H}$
is the total energy of the (strained) Fe-H system. It can be rewritten in terms of the solution
energy $W$ (\textit{c.f.} Eq.~\ref{solW}) as $E_{Fe-H}=E_{Fe}+W-\mu n$, where 
we have set the energy per H as $-\mu+\frac{1}{2}E_{H_2}$, in contrast
to Eq.~\ref{solW}, where it was set to $\frac{1}{2}E_{H_2}$.

The free energy is minimized with respect to $n$:
\begin{equation}
\frac{\delta F}{\delta n}=0=\frac{\partial W}{\partial n}-T\frac{\partial S}{\partial n}-\mu.
\label{omega_d}
\end{equation}
We first focus on the case of a hydrostatic strain field. 
For simplicity we truncate Equation~\ref{hydroeqn} from the previous section to linear order in strain: 
\begin{equation}
W\equiv W(\epsilon)=n(\Delta E-3B^0\Delta V\epsilon).
\label{Wsimple}
\end{equation}
The expression for the entropy can take on various forms, in the presence of
various types of H-H interactions, all equivalent at low concentrations.
In what follows, $n$ is the number of H, $N$ is the number of matrix (Fe) atoms, $M=6N$ is the total
number of tetrahedral interstitial lattice sites, and $c$ is the occupation concentration defined
as $c=n/M$.
The simplest form for the entropy, $S_0$, arises from an unrestricted concentration of H, 
corresponding not to a lattice but to an ideal gas. 
In this case, the entropy derivative is 
$\frac{\partial S_0}{\partial n}=-k \mathrm{ln}\frac{n}{M}$. The concentration yielded by Eq.~\ref{omega_d}
for the case of hydrostatic strain up to linear order in the strain 
takes the Maxwell-Boltzmann form of 
\begin{equation}
c=c_0\mathrm{exp}\left[3B^0\Delta V\epsilon/kT\right]
\end{equation}
where $c_0\equiv\mathrm{exp}\left[(-\Delta E+\mu)/kT\right]$. This reproduces the result from the Cottrell picture
of enhanced concentration around a dislocation~\cite{Cottrell}.
 A more realistic form of the entropy takes into account
the presence of a lattice and enforces a maximum occupancy of one H per tetrahedral site.
In this case the entropy is $S_1=-k\left[n\mathrm{ln}\frac{n}{M}+(M-n)\mathrm{ln}\left(1-\frac{n}{M}\right)\right]$
and the entropy derivative is $\frac{\partial S_1}{\partial n}=-k \mathrm{ln}\frac{n}{M-n}$. The concentration using
this form follows Fermi-Dirac statistics:
\begin{equation}
c=\left\{\mathrm{exp}\left[\left(\Delta E-\mu-3B^0\Delta V\epsilon\right)/kT\right]+1\right\}^{-1}.
\end{equation}
However, the maximum occupancy of one H per tetrahedral site is not realistic as the H-H interaction 
in Fe is repulsive: We have calculated the H-H interaction as a function of distance 
and find strong H-H repulsion up to third-nearest neighbours where it was 0.13 eV. 
 For a realistic representation of the statistics, up to and including third nearest-neighbour H-H spacings
should be strongly unfavourable in our model. In the extreme case, these sites are left unoccupied, corresponding to
infinite repulsion. The expression
for the entropy derivative 
for tetrahedral interstitials in a bcc lattice with total exclusion of first nearest-neighbours, both first and second-nearest-neighbours,
and up to third-nearest neighbours 
is given in Ref.~\cite{boureau} as 
\begin{eqnarray}
\frac{\partial S_{1nn}}{\partial n}&=&-k\mathrm{ln}\frac{6n}{z(n+Mz)};\qquad z=\frac{(n-3M)(5n^2-36nM+72M^2)}{(n-6M)^2M},\nonumber\\
\frac{\partial S_{2nn}}{\partial n}&=&-k\mathrm{ln}\frac{n(M-n)}{(M-4n)^2},\nonumber\\
\frac{\partial S_{3nn}}{\partial n}&=&-k\mathrm{ln}\left[\frac{n(M-3n)(M-2n)^3}{(M-4n)(M-5n)^4}\right]
\end{eqnarray}
respectively. In the limit of low concentration, all of the expressions 
for the entropy ($S_0$, $S_1$, $S_{1nn}$, $S_{2nn}$, $S_{3nn}$)
 are the same, the differences
being of $O(c^2)$. Similarly, at small strains, all of the solutions for $c$ are identical. Therefore,
the H-H interaction can be ignored for predicting H concentrations around dislocations.

\begin{figure}[htp]
\includegraphics[angle=0,width=5.5in]{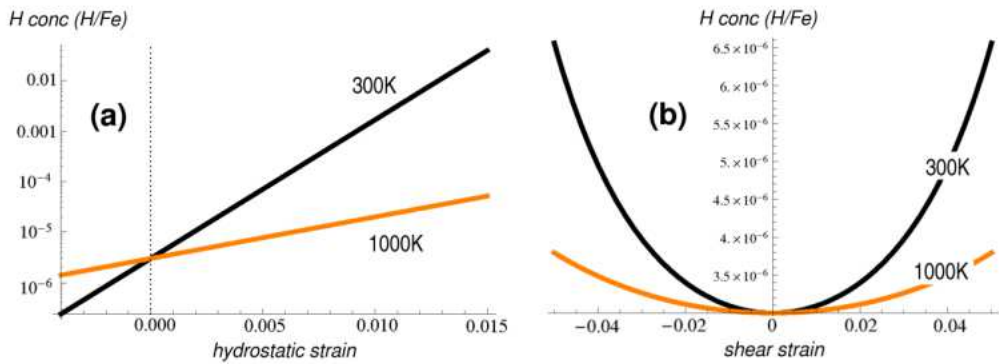}
\caption{[color online] Concentration of H as a function of (a) hydrostatic strain, and (b) shear strain 
$\epsilon_{\theta^\prime z^\prime}$ where the dislocation line $z^\prime$ is taken as the $\langle 111 \rangle$ 
direction, for a chemical potential giving a background concentration of $3\times 10^{-6}$ H per Fe atom
at each temperature. The case of constant volumes for Fe and Fe-H was assumed.}
\label{conc_low}
\end{figure}

A numerical example of the extent of the modification of the H-concentration in bcc Fe by 
applied hydrostatic strain is shown in 
Figure~\ref{conc_low}a. In this low-concentration regime, all forms of the entropy are identical. 
In this example, the background (zero-strain) atomic H concentration $\left( n/N\right)$ was 
taken to be equal to 3$\times 10^{-6}$,
in line with typical densities of H in iron~\cite{Hirth}. At each temperature studied, $\mu$ was adjusted 
in order to achieve this background concentration. From the figure, it is clear that the H-concentration can be
dramatically increased by orders of magnitude, even at very small strains. The general 
trend of increased H concentration under tensile strain supports the interpretation 
of H solubility as a volume effect. The increase in concentration is 
more pronounced at lower temperatures, in agreement with the low thermal desoprtion 
temperatures, \textit{ca} 400 K, indicated for H trapped at 
dislocations~\cite{Novak}. We take our results of Fig.~\ref{conc_low}a to be 
representative of edge dislocations which have a significant hydrostatic strain 
component~\cite{HirthLothe}. With shear strain applied 
in the ${\theta^\prime z^\prime}$ direction where $z^\prime$ is the 
$\langle 111 \rangle$ direction, and regardless of the direction of $\theta^\prime$, we obtain the 
curves in Fig.~\ref{conc_low}b. The 
effect is less pronounced than for hydrostatic strain, because
the solubility energy is quadratic in the shear strain 
(\textit{c.f.} Eq.~\ref{pshreqn}, whose coefficients change upon
rotation to this coordinate system, but whose functional form remains the same).

\begin{figure}[htp]
\includegraphics[angle=0,width=5.5in]{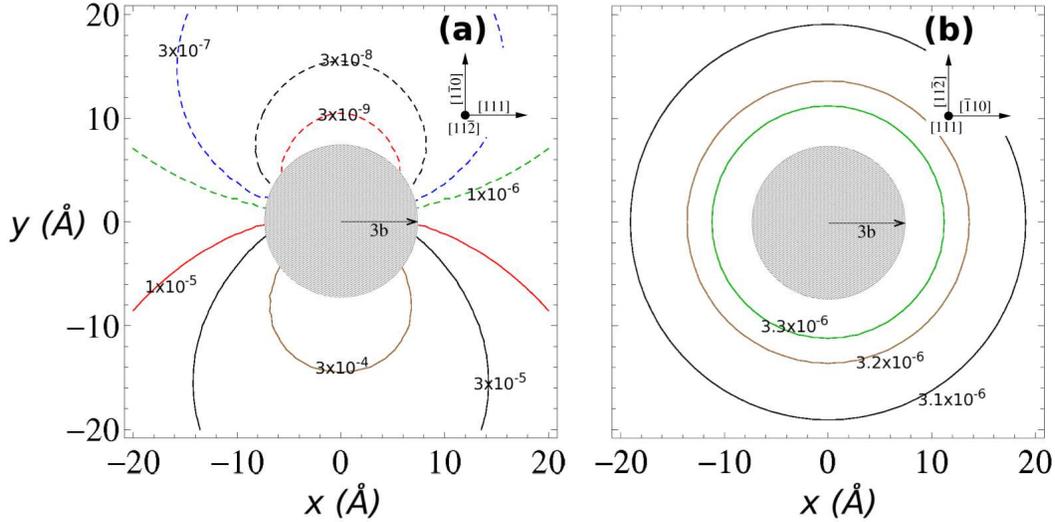}
\caption{[color online] Concentration contours of H around an (a) edge dislocation, with the tensile region at $y<0$ 
and (b) screw dislocation, at 300 K, for a background concentration of $3\times 10^{-6}$ H per Fe atom. The case of constant volumes
for Fe and Fe-H was assumed. The shaded circle denotes a
region of radius three times the Burgers vector at the core.}
\label{conc_edge-screw}
\end{figure}

By combining component strains in our solubility expressions, we are able to predict H concentration profiles
in the full strain fields of edge, as well as screw dislocations. H concentration profiles are shown for two common slip systems in $\alpha$-Fe in 
Fig.~\ref{conc_edge-screw}. It is expected that symmetry reduction will result from the consideration of tetragonal distortions, as for the case of
C in Fe in Ref.~\cite{chrzan}, and by the inclusion of dislocation anisotropy~\cite{HirthLothe,Clouet11}.

\section{Conclusions}
\label{sec:conclusions}
We have derived a model for the H point-defect - dislocation interaction (solubility) 
in $\alpha$-Fe and parametrised it with \textit{ab initio}-derived 
 quantities from a previous study~\cite{psiachos1}. Although, for simplicity,
our model assumes a hydrostatic lattice expansion due to H, we find not only an effect on the solubility
from axial strain, but from shear strain as well. In order to get an interaction with screw dislocations, it
was necessary to go beyond the H-induced size effect and consider the modulus effect~\cite{eshelby1951,fiore,HirthLothe,wolfer}.
From our models we find, in agreement with existing experimental studies~\cite{Ono,Novak}, and a recent
atomistic study~\cite{chrzan}, that typical strain fields 
found near dislocations can trap substantial amounts of H. We further 
show that the local H-concentration may be enhanced by several
orders of magnitude in hydrostatic strain fields, observed in edge dislocations, 
and this enhancement identifies a realistic regime for the consideration of H concentrations in the atomic \% range
as in our previous study~\cite{psiachos1}. In the case of shear strain, which 
constitutes the strain field of screw dislocations,
despite our simplifications of hydrostatic expansion of the lattice by H, we also find attraction 
of H if the screw dislocation is oriented along the $\langle 111\rangle$ direction. However,
 the concentration enhancement for shear strain is not as great as 
in the case of hydrostatic strain.
We find, in general, that the areas of enriched local
concentration tend to occur mainly at low temperatures, in the hundreds of K. Similar conclusions regarding
the enrichment of C in $\alpha$-iron as a function of strain and temperature were reached in a recent 
atomistic study~\cite{chrzan}. Screw 
dislocations along $\langle 111\rangle$
 are the predominant types of dislocations in bcc metals driving plasticity~\cite{screw1} and
the concentration enhancement which we find near screw dislocations 
is consistent with various theories 
and observations of H trapping at dislocations. Our findings support the 
strong role played by dislocations~\cite{Ono,Novak} in describing the effect of 
H on the mechanical properties bcc Fe.

\section*{Acknowledgements}
Discussions with Thomas Hammerschmidt are gratefully acknowledged.



\vspace{6mm}
\begin{thebibliography}{10}

\bibitem{Oriani}
R.~A. Oriani.
\newblock Hydrogen--the versatile embrittler.
\newblock {\em Corros.}, 43:390--397, 1987.

\bibitem{Hirth}
J.~P. Hirth.
\newblock Effects of hydrogen on the properties of iron and steel.
\newblock {\em Metall. Mater. Trans. A}, 11:861--890, 1980.

\bibitem{Troiano}
A.~R. Troiano.
\newblock The role of hydrogen and other interstitials on the mechanical
  behavior of metals.
\newblock {\em Trans. Am. Soc. Met.}, 52:54--80, 1960.

\bibitem{Oriani2}
R.~A. Oriani and P.~H. Josephic.
\newblock Equilibrium and kinetic studies of the hydrogen-assisted cracking of
  steel.
\newblock {\em Acta Metall.}, 25:979--988, 1977.

\bibitem{DawBaskes}
M.~S. Daw and M.~I. Baskes.
\newblock Semiempirical, quantum mechanical calculation of hydrogen
  embrittlement in metals.
\newblock {\em Phys. Rev. Lett.}, 50:1285--1288, 1983.

\bibitem{Ohno}
Y.~Tateyama and T.~Ohno.
\newblock Atomic-scale effects of hydrogen in iron toward hydrogen
  embrittlement: Ab-initio study.
\newblock {\em ISIJ Int.}, 43:573--578, 2003.

\bibitem{TateyamaPRB}
Y.~Tateyama and T.~Ohno.
\newblock Stability and clusterization of hydrogen-vacancy complexes in
  $\alpha$-Fe: An ab initio study.
\newblock {\em Phys. Rev. B}, 67:174105--1--10, 2003.

\bibitem{Beachem}
C.~D. Beachem.
\newblock A new model for hydrogen-assisted cracking (hydrogen
  “embrittlement”).
\newblock {\em Metall. Mater. Trans. B}, 3:441--455, 1972.

\bibitem{BirnbaumMatSciEng}
H.~K. Birnbaum and P.~Sofronis.
\newblock Hydrogen-enhanced localized plasticity - a mechanism for
  hydrogen-related fracture.
\newblock {\em Mater. Sci. Eng. A}, 176:191--202, 1994.

\bibitem{Robertson}
I.~M. Robertson.
\newblock The effect of hydrogen on dislocation dynamics.
\newblock {\em Eng. Fract. Mech.}, 68:671--692, 2001.

\bibitem{Cottrell}
A.~H. Cottrell.
\newblock Effects of solute atoms on the behaviour of dislocations.
\newblock In {\em Report of a conference on the strength of solids}, pp~30-36, 1948. The Physical Society, London;
A.~H. Cottrell and B.~A. Bilby.
\newblock Dislocation theory of yielding and strain ageing of iron.
\newblock {\em Proc. Phys. Soc.}, 62:49--61, 1949.

\bibitem{Kirchheim}
R.~Kirchheim.
\newblock Revisiting hydrogen embrittlement models and hydrogen-induced
  homogeneous nucleation of dislocations.
\newblock {\em Scr. Mater.}, 62:67--70, 2010.

\bibitem{hayashi}
Y.~Hayashi and W.~M. Shu.
\newblock Iron (ruthenium and osmium)-hydrogen systems.
\newblock {\em Solid State Phenom.}, 73-75:65--114, 2000.

\bibitem{JiangCarter}
D.~E. Jiang and E.~A. Carter.
\newblock Diffusion of interstitial hydrogen into and through bcc Fe from first
  principles.
\newblock {\em Phys. Rev. B}, 70:064102--1--9, 2004.

\bibitem{Ono}
K.~Ono and M.~Meshii.
\newblock Hydrogen detrapping from grain boundaries and dislocations in high
  purity iron.
\newblock {\em Acta Metall. Mater.}, 40:1357--1364, 1992.

\bibitem{matsui}
H.~Matsui, H.~Kimura, and S.~Moriya.
\newblock The effect of hydrogen on the mechanical properties of high purity
  iron i. softening and hardening of high purity iron by hydrogen charging
  during tensile deformation.
\newblock {\em Mater. Sci. Eng.}, 40:207--216, 1979.

\bibitem{murakami}
Y.~Murakami, T.~Kanezaki, and Y.~Mine.
\newblock Hydrogen effect against hydrogen embrittlement.
\newblock {\em Metall. Mater. Trans. A}, 41:2548--2562, 2010.

\bibitem{Novak}
P.~Novak, R.~Yuan, B.~P. Somerday, P.~Sofronis, and R.~O. Ritchie.
\newblock A statistical, physical-based, micro-mechanical model of
  hydrogen-induced intergranular fracture in steel.
\newblock {\em J. Mech. Phys. Solids}, 58:206--226, 2010.

\bibitem{Gerberich}
W.~W. Gerberich, D.~D. Stauffer, and P.~Sofronis.
\newblock A coexistent view of hydrogen effects on mechanical behavior of
  crystals: HELP and HEDE.
\newblock In B.~Somerday, P.~Sofronis, and R.~Jones, editors, {\em Effects of
  Hydrogen on Materials}, pp~38-45, Materials Park OH, 2009. ASM International.

\bibitem{Taketomi2}
S.~Taketomi, R.~Matsumoto, and N.~Miyazaki.
\newblock Atomistic study of hydrogen distribution and diffusion around a
  \{112\}$\langle 111\rangle$ edge dislocation in alpha iron.
\newblock {\em Acta Mater.}, 56:3761--3769, 2008.

\bibitem{Rama}
A.~Ramasubramaniam, M.~Itakura, M.~Ortiz, and E.~A. Carter.
\newblock Effect of atomic scale plasticity on hydrogen diffusion in iron:
  Quantum mechanically informed and on-the-fly kinetic Monte Carlo simulations.
\newblock {\em J. Mater. Res.}, 23:2757--2773, 2008.

\bibitem{ClouetActaMat}
E.~Clouet, S.~Garruchet, H.~Nguyen, M.~Perez, and C.~S. Becquart.
\newblock Dislocation interaction with C in $\alpha$-Fe: A comparison between
  atomistic calculations and elasticity theory.
\newblock {\em Acta Mater.}, 56:3450--3460, 2008.

\bibitem{Rama2}
A.~Ramasubramaniam, M.~Itakura, and E.~A. Carter.
\newblock Interatomic potentials for hydrogen in α–iron based on density
  functional theory.
\newblock {\em Phys. Rev. B}, 79:174101--1--13, 2009.

\bibitem{Sanchez}
J.~Sanchez, J.~Fullea, C.~Andrade, and P.~L. de~Andres.
\newblock Hydrogen in $\alpha$-iron: Stress and diffusion.
\newblock {\em Phys. Rev. B}, 78:014113--1--7, 2008.

\bibitem{SanchezMD}
J.~Sanchez, J.~Fullea, M.~C. Andrade, and P.~L. de~Andres.
\newblock Ab initio molecular dynamics simulation of hydrogen diffusion in
  $\alpha$-iron.
\newblock {\em Phys. Rev. B}, 81:132102--1--4, 2010.

\bibitem{BecquartHeW}
C.~S. Becquart, C.~Domain, U.~Sarkar, A.~DeBacker, and M.~Hou.
\newblock Microstructural evolution of irradiated tungsten: Ab initio
  parameterisation of an OKMC model.
\newblock {\em J. Nucl. Mater.}, 403:75--88, 2010.

\bibitem{sluiter}
S.~Vannarat, M.~H.~F. Sluiter, and Y.~Kawazoe.
\newblock First-principles study of solute-dislocation interaction in
  aluminum-rich alloys.
\newblock {\em Phys. Rev. B}, 64:224203--1--8, 2001.

\bibitem{Gavriljuk}
V.~G. Gavriljuk, V.~N. Shivanyuk, and B.~D. Shanina.
\newblock Change in the electron structure caused by C, N and H atoms in iron
  and its effect on their interaction with dislocations.
\newblock {\em Acta Mater.}, 53:5017--5024, 2005.

\bibitem{cochardt}
A.~W. Cochardt, G.~Schoek, and H.~Wiedersich.
\newblock Interaction between dislocations and interstitial atoms in
  body-centered cubic metals.
\newblock {\em Acta Metall.}, 3:533--537, 1955.

\bibitem{eshelby1957}
J.~D. Eshelby.
\newblock The determination of the elastic field of an ellipsoidal inclusion,
  and related problems.
\newblock {\em Proc. R. Soc. Lond. A}, 241:376--396, 1957.

\bibitem{bilby}
B.~A. Bilby.
\newblock On the interactions of dislocations and solute atoms.
\newblock {\em Proc. Phys. Soc.}, 63:191--200, 1950.

\bibitem{Hirth2}
J.~P. Hirth and B.~Carnahan.
\newblock Hydrogen adsorption at dislocations and cracks in Fe.
\newblock {\em Acta Metall.}, 26:1795--1803, 1978.

\bibitem{bockris}
J.~O'M. Bockris, W.~Beck, M.~A. Genshaw, P.~K. Subramanyan, and F.~S. Williams.
\newblock The effect of stress on the chemical potential of hydrogen in iron
  and steel.
\newblock {\em Acta Metall.}, 19:1209, 1971.

\bibitem{Zhang_strain}
T.-Y. Zhang, F.-X. Jiang, W.-Y. Chu, and C.-M. Hsiao.
\newblock Tetragonal distortion field of hydrogen atoms in iron.
\newblock {\em Metall. Mater. Trans. A}, 16:1649--1653, 1985.

\bibitem{fiore}
N.~F. Fiore and C.~L. Bauer.
\newblock Binding of solute atoms to dislocations.
\newblock {\em Prog. Mater. Sci.}, 13:85--134, 1967.

\bibitem{chrzan}
Y.~Hanlumyuang, P.~A. Gordon, T.~Neeraj, and D.~C. Chrzan.
\newblock Interactions between carbon solutes and dislocations in bcc iron.
\newblock {\em Acta Mater.}, 58:5481--5490, 2010.

\bibitem{eshelby1951}
J.~D. Eshelby.
\newblock The force on an elastic singularity.
\newblock {\em Philos. Trans. R. Soc. A}, 244:87--112, 1951.

\bibitem{HirthLothe}
J.~P. Hirth and J.~Lothe.
\newblock {\em Theory of Dislocations}.
\newblock Krieger Pub. Co., 2nd edition, 1992.

\bibitem{wolfer}
W.~G. Wolfer.
\newblock The dislocation bias.
\newblock {\em J. Comput.-Aided Mater. Des.}, 14:403--417, 2007.

\bibitem{psiachos1}
D.~Psiachos, T.~Hammerschmidt, and R.~Drautz.
\newblock Ab initio study of the modification of elastic properties of $\alpha$-iron
  by hydrostatic strain and by hydrogen interstitials.
\newblock {\em Acta Mater.}, 59:4255--4263, 2011.

\bibitem{vasp1}
G.~Kresse and J.~Hafner.
\newblock Ab initio molecular dynamics for open-shell transition metals.
\newblock {\em Phys. Rev. B}, 48:13115--13118, 1993.

\bibitem{vasp2}
G.~Kresse and J.~Furthm\"{u}ller.
\newblock Efficiency of ab-initio total energy calculations for metals and
  semiconductors using a plane-wave basis set.
\newblock {\em Comput. Mater. Sci.}, 6:15--50, 1996.

\bibitem{vasp3}
G.~Kresse and J.~Furthm\"{u}ller.
\newblock Efficient iterative schemes for ab initio total-energy calculations
  using a plane-wave basis set.
\newblock {\em Phys. Rev. B}, 54:11169--11186, 1996.

\bibitem{paw1}
P.~Bl\"{o}chl.
\newblock Projector augmented-wave method.
\newblock {\em Phys. Rev. B}, 50:17953--17979, 1994.

\bibitem{paw2}
G.~Kresse and D.~Joubert.
\newblock From ultrasoft pseudopotentials to the projector augmented-wave
  method.
\newblock {\em Phys. Rev. B}, 59:1758--1775, 1999.

\bibitem{pw91}
J.~P. Perdew and Y.~Wang.
\newblock Accurate and simple analytic representation of the electron-gas
  correlation energy.
\newblock {\em Phys. Rev. B}, 45:13244--13249, 1992.

\bibitem{LiChen1}
D.~Y. Li and L.~Q. Chen.
\newblock Shape of a rhombohedral coherent Ti$_{11}$Ni$_{14}$ precipitate in a
  cubic matrix and its growth and dissolution during constrained aging.
\newblock {\em Acta Mater.}, 45:2435--2442, 1997;
D.~Y. Li and L.~Q. Chen.
\newblock Morphological evolution of coherent multi-variant Ti$_{11}$Ni$_{14}$
  precipitates in ni-ti alloys under an applied stress-a computer simulation
  study.
\newblock {\em Acta Mater.}, 46:639--649, 1998.

\bibitem{Ryu}
S.~Ryu and K.~Kang and W.~Cai.
\newblock Entropic effect on the rate of dislocation nucleation.
\newblock {\em Proc. Natl. Acad. Sci. USA}, 108:5174--5178, 2011.

\bibitem{barron}
T.~H.~K. Barron and M.~L. Klein.
\newblock Second-order elastic constants of a solid under stress.
\newblock {\em Proc. Phys. Soc.}, 85:523--532, 1965.

\bibitem{Wallace}
D.~C. Wallace.
\newblock {\em Thermodynamics of Crystals}.
\newblock Dover, 1972.

\bibitem{eshelbySSP}
J.~D. Eshelby.
\newblock The continuum theory of lattice defects.
\newblock {\em Solid State Phys.}, 3:79--144, 1956.

\bibitem{screw1}
K.~Ito and V.~Vitek.
\newblock Atomistic study of non-Schmid effects in the plastic yielding of bcc
  metals.
\newblock {\em Philos. Mag. A}, 81:1387--1407, 2001.

\bibitem{Taketomi3}
R.~Matsumoto, Y.~Inoue, S.~Taketomi, and N.~Miyazaki.
\newblock Influence of shear strain on the hydrogen trapped in bcc-Fe: A
  first-principles-based study.
\newblock {\em Scr. Mater.}, 60:555--558, 2009.

\bibitem{SofronisJMechPhysSol}
P.~Sofronis and H.~K. Birnbaum.
\newblock Mechanics of the hydrogen-dislocation-impurity interactions—i.
  increasing shear modulus.
\newblock {\em J. Mech. Phys. Solids}, 43:49--90, 1995.

\bibitem{boureau}
G.~Boureau.
\newblock A simple method of calculation of the configurational entropy for
  interstitial solutions with short range repulsive interactions.
\newblock {\em J. Phys. Chem. Solids}, 42:743--748, 1981.

\bibitem{Clouet11}
R.~G.~A. Veiga, M.~Perez, C.~S. Becquart, E.~Clouet, and C.~Domain.
\newblock Comparison of atomistic and elasticity approaches for carbon
  diffusion near line defects in $\alpha$-iron.
\newblock {\em Acta Mater.}, 59:6963--6974, 2011.

\end{thebibliography}

\end{document}